\documentclass[%
 aps,
rsi,%
 amsmath,amssymb,
 reprint,%
]{revtex4-1}

\usepackage{comment}
\usepackage{graphicx}% Include figure files
\usepackage{dcolumn}% Align table columns on decimal point
\usepackage{bm}% bold math
\usepackage{afterpage}

\newcommand{\etal}{\textit{et al}.}
\newcommand{\ie}{\textit{i}.\textit{e}.}

\usepackage{lipsum}

\begin{document}

\preprint{AIP/123-QED}

\title[]{Structure of Multicomponent Coulomb Crystals}% Force line breaks with \\
%\thanks{Footnote to title of article.}

\author{M. E. Caplan}
% \altaffiliation[Also at ]{Physics Department, XYZ University.}%Lines break automatically or can be forced with \\
%\author{B. Author}%
 \email{mecapl1@ilstu.edu}
\affiliation{Illinois State University, Normal, IL 61761 USA}

%\author{C. Author}
% \homepage{http://www.Second.institution.edu/~Charlie.Author.}
%\affiliation{%
%Second institution and/or address%\\This line break forced% with \\
%}

\date{\today}% It is always \today, tolowday,
             %  but any date may be explicitly specified

\begin{abstract}
Coulomb plasmas crystallize in a number of physical systems, such as dusty plasmas, neutron star crusts, and white dwarf cores. The crystal structure of the one component and binary plasma has received significant attention in the literature, though the less studied multicomponent plasma may be most relevant for many physical systems which contain a large range of particle charges. We report on molecular dynamics simulations of multicomponent plasmas near the melting temperature with mixtures taken to be realistic X-ray burst ash compositions. We quantify the structure of the crystal with the bond order parameters and radial distribution function. Consistent with past work, low charge particles form interstitial defects and we argue that they are in a quasi-liquid state within the lattice. The lattice shows screening effects which preserves long range order despite the large variance in particle charges, which may impact transport properties relevant to astrophysics. %We argue that the usual assumption of randomly distributed ions in the impurity parameter formalism results in an overestimate of crystal disorder and resistivity.

\end{abstract}

\pacs{Valid PACS appear here}% PACS, the Physics and Astronomy
                             % Classification Scheme.
\keywords{Molecular dynamics, bond order parameter, Coulomb plasma}%Use showkeys class option if keyword
                              %display desired
\maketitle

\section{Introduction}\label{sec:intr}

Coulomb (Yukawa) plasmas consist of a set of charged point particles interacting via a Coulomb repulsion which is screened by a neutralizing background gas. At sufficiently high pressure (or density) these systems can crystallize even despite astronomically high temperatures. 

%Terrestrial coulomb plasma experiments, such as dusty plasmas studied in microgravity, have been studied as analogs for Coulomb plasmas in stars. 
The properties of these `astromaterials,' materials present in stars, impact observations relevant to a number of astrophysical phenomena. To name a few, (1) latent heat release in freezing white dwarfs is now observed to affect the cooling of white dwarfs \cite{Tremblay2019}, (2) phase separation of heavy nuclei in the oceans of freezing white dwarfs may release additional gravitational potential energy which affects cooling \cite{Tremblay2019,Bildsten2001}, and (3) the thermal transport properties of the phase separated neutron star crust in accreting X-ray binaries must be known to interpret observations of X-ray binaries in quiescence \cite{Caplan2018,CaplanRMP,Brown_2009}. 

The one-component plasma (OCP) is now well studied and is known to undergo a first order phase transition between a liquid and solid phase \cite{Vaulina2002,BAUS19801,CaplanRMP}. Phase separation and diffusion of the binary mixture has also received attention in the literature \cite{Daligault2012,Shaffer2017,Kozhberov2015}. Analytic models of ternary mixtures have been used to approximate the phase separation of mixtures with many components near their eutectic point \cite{Caplan2018}. Past work studying the structure of a few specific multicomponent plasma (MCP) mixtures found that they tended to form a bcc lattice with a number of complicated compositionally driven defects. \cite{HorowitzNSC,Daligault2009,Potekhin2009,Horowitz2007,Horowitz2009,Caplan2018}. 

%Horowitz \etal\ calculated the static structure factor for an MCP and reports a thermal conductivity for accreted neutron star crusts \cite{Horowitz2009} 

Electron scattering in MCPs is relevant for astrophysics, as electron-impurity scattering affects the thermal and electrical conductivity (see for example \citealt{Brown_2009}). 
Astrophysical models use the impurity parameter,
$Q_{imp} = (1 / n) \Sigma_i n_i ( \bar{Z} - Z_i)^2 $ (defined as the variance in mixture charge), to calculate mixture transport properties. However, the impurity parameter formalism may overpredict the electron scattering frequency by assuming a random distribution of impurities \cite{Ogata1993,Itoh1998}. While this assumption may be valid for nearly pure mixtures with trace impurities, mixtures with many components may have more complicated lattice structures which may not necessarily enhance the electron scattering rate as much as a random impurity distribution might predict. For example, consider a binary ionic mixture (BIM) studied by \citealt{Ogata1993}. A BIM with equal abundances of species $Z_1$ and $Z_2$ may have a ground state CsCl-structure, with each ion of species 1 being at the center of a cube of eight ions of species 2 and vice versa. With this long range order such a crystal would effectively have no defects, though a large $Z_1 /Z_2$ would yield a high $Q_{imp}$ suggesting otherwise.

Simulations studying specific MCPs are thus well motivated astrophysically. In past work, Horowitz \etal\ studied the phase separation of one mixture and calculated the static structure factor $S(q)$ of the resulting solid to determine the thermal transport properties \cite{Horowitz2007,Horowitz2009}. \citealt{Roggero2016} have recently used a Path Integral Monte Carlo approach to determine the electron scattering rate in multicomponent plasmas and find that the effective lattice impurity is reduced by a factor of a few relative to theoretical calculations using the impurity parameter.  In contrast, having a large number of species may introduce many new kinds of disorder (relative to the OCP) even if the lattice structure is regular as in the BIM case considered above. Trace low $Z$ impurities have been observed to form clusters of interstitial defects, suggesting they separate and form pockets of disorder within the lattice \cite{Caplan2018,Horowitz2009}. Furthermore, phase separation may produce crystalline domains which are locally purified in order to accommodate the entire mixture. Grain boundaries between such `compositional domains' could act as sites for electron scattering which may be important depending on their size \cite{Caplan2018}.

%In this work we seek to understand this reduction in terms of the physical lattice structure, and to study if this effect can be understood more generally for arbitrary mixtures. As the physical properties of MCPs are often calculated from linear mixing theory, we seek to we seek to resolve nonlinear effects in the lattice structure with molecular dynamics simulations.

Beyond the astrophysical implications, results from simulations of MCPs may be generalized to the study of dusty plasmas, an analogous system where macroscopic charged particulates are allowed to interact via the Coulomb force (often in zero gravity) which is well studied in terrestrial experiments \cite{CaplanRMP,RevModPhys.81.25,RevModPhys.81.1353,FORTOV20051}. The structure of two-dimensional crystalline dusty plasmas is well studied, particularly for binary mixtures, and shows strong agreement with theoretical predictions such as those from MD simulations (especially for properties such as crystal structure, diffusion, and phase separation) \cite{Zonko2009,PhysRevE.58.7831}. Diffusion of dust particles in the lattice %, likewise of interest for astrophysical MD simulations, 
is experimentally accessible with high speed cameras and fluorescent techniques \cite{PhysRevLett.116.115002}. These phenomena are both relevant to astrophysics and easily accessible with MD. Therefore, simulations studying astrophysically motivated mixtures may motivate terrestrial dusty plasma experiments if systems with similar charge mixtures can be produced.

In this work we study the structure of six crystalline MCPs produced from molecular dynamics (MD) simulations using the bond order parameter and radial distribution functions. In Sec. \ref{sec:for} we discuss our formalism, including our MD code (Sec \ref{sec:md}), our methods of calculating the bond order parameter and radial distribution functions (Secs. \ref{ss:bop} and \ref{ss:gr}), and a brief discussion of the theory of OCPs and MCPs (Sec. \ref{ss:cp}). We describe our simulations and results in Sec. \ref{sec:sim} and conclude in Sec. \ref{sec:dis}.

\section{\label{sec:for}Formalism}

%In this section we discuss our molecular dynamics formalism, and our computational methods for characterizing the structure of the crystal.

\subsection{\label{sec:md}Molecular Dynamics}

Nuclei in Coulomb plasmas are fully ionized and are treated as point particles (ions) which interact via a two-body screened Coulomb potential 

\begin{equation}
V(r_{ij})=\frac{Z_i Z_j e^2}{r_{ij}} \exp(-r_{ij}/\lambda),
\label{eq.V}
\end{equation}  

\noindent where $Z_i$ and $Z_j$ are the electric charges of the $i^{\text{th}}$ and $j^{\text{th}}$ nuclei and $r_{ij}$ is the separation between them. The exponential term is due to the screening from the degenerate electron gas between ions and is calculated using the Thomas Fermi screening length $\lambda^{-1}=2\alpha^{1/2}k_F/\pi^{1/2}$ using the fine structure constant $\alpha$ and electron Fermi momentum $k_F=(3\pi^2n_e)^{1/3}$. For the MCP we require the electron density $n_e$ to be equal to the charge density from the ions (\textit{i.e.} $n_e=\langle Z\rangle n$ with ion number density $n$ and average charge $\langle Z\rangle$). Electrons are not included explicitly; their effects on the lattice are included through the potential screening.

To evolve the system we solve Newton's equations of motion numerically using a velocity Verlet scheme with the Indiana University Molecular Dynamics (IUMD) CUDA-Fortran code, version 6.3.1. This code has been used extensively to study astromaterials in neutron star crusts and white dwarfs and is described in more detail in past work \cite{Caplan2018,Horowitz2009,HorowitzNSC}. All simulations presented in this work use periodic boundary conditions and cubic simulation volumes. 

\subsection{Bond Order Parameter}\label{ss:bop}

We quantify the local order of the lattice around nuclei of each species in our simulations with the bond order parameter $Q_6$. This allows us to evaluate the `solidness' or `liquidness' of each nucleus with a simple metric determined from the relative positions of its nearest neighbors. For an individual ion, $Q_6$ generally takes on a value between 0 and 0.5. Past work by \citealt{Lechner2008} have shown that in Lennard-Jones mixtures at finite temperatures one expects $Q_6 \sim 0.44$ in a bcc lattice, which is similar to what was reported by \citealt{Caplan2018} when studying phase separation in Coulomb crystals. %0806.3345 

We calculate $Q_6$ (as in \citealt{Wang2005}) by

\begin{equation}
Q_6 = \sqrt{ \frac{4 \pi}{13} \sum_{m=-6}^{6} \left | \frac{1}{N_b} \sum_{\rm bonds} Y_{6m} ( \theta ( \mathbf{r} ) , \phi ( \mathbf{r} ) ) \right |^2 } 
\end{equation}

\noindent for each particle. Spherical harmonics $Y_{6m}$ are calculated from the angles $\theta(\mathbf{r})$ and $\phi(\mathbf{r})$ of the vector between pairs of nuclei. This is averaged over nearest neighbors and harmonics to produce the coordinate independent $Q_6$. 

\subsection{Radial Distribution Function}\label{ss:gr}

We calculate the radial distribution function $g(r)$ for ions in our mixtures. As mixtures contain many species, we must consider the radial distribution functions between species $g_{ij}(r)$, following the definition used by \citealt{Thorneywork},

\begin{equation}
c_i c_j g_{ij}(r) = \frac{1}{N \rho}  \langle \sum_{\mu=1}^{N_i} \sum_{\nu \neq \mu }^{N_j}  \delta( \bf{r} +  \bf{r}_\mu - \bf{r}_\nu  ) \rangle
\end{equation}

\noindent where $c_{i,j}$ are the concentrations of the $i^{th}$ and $j^{th}$ species, $N$ is the total number of particles in the mixture, $\rho$ is the total number density, and the sums are over all pairs of particles of species $i$ and $j$. Distances $\bf{r}_\mu - \bf{r}_\nu$ are taken over the periodic boundary.
For an OCP we recover the radial distribution function for $i=j$. For simplicity and due to the large number of species in our mixtures we will only consider $i$ as the most abundant species in the mixture (for all of our mixtures $c_i \gtrsim 0.25$). Our normalizations are such that $g_{ij}(r) = 1$ as $r\to\infty$.

\subsection{Coulomb Plasmas}\label{ss:cp}

\subsubsection{One-component Plasmas}\label{sss:ocp}

We discuss one-component plasmas as they will be used for reference when studying mixtures. The one component plasma is characterized by the Coulomb plasma parameter $\Gamma$, calculated as

\begin{equation}\label{eq:gamma}
    \Gamma_{OCP} = \frac{e^2 Z^2}{a T}
\end{equation}

\noindent with squared elementary charge $e^2$ ($\approx 1.44$ MeV fm), ion charge $Z$, Wigner-Seitz radius $a = (4 \pi n / 3)^{-1/3}$ with $n$ ion number density as before, and temperature $T$ (in MeV). The critical $\Gamma_{crit} = 175$ occurs at the melting temperature. It is often useful to report $\Gamma_{crit}/ \Gamma$, which is linear with temperature and is unity at the melting temperature. When $\Gamma_{crit}/\Gamma < 1$, the OCP can form a bcc or fcc lattice, though the bcc lattice is typically the relevant case for neutron stars \cite{Vaulina2002,CaplanRMP}. 

\subsubsection{Multi-component Plasmas}\label{sss:mcp}

In a mixture, each component of charge $Z_i$ and concentration $c_i$ can be characterized individually by $\Gamma_i = e^2 Z_i^2 / a_i T$ where $a_i$ must now be defined in terms of the average charge density of the mixture $\rho_{ch}$, $a_i = (3Z_i / 4 \pi \rho_{ch})^{1/3}$. Averaging over all components gives

\begin{equation}
\Gamma_{MCP} = \frac{ \langle Z^{5/3} \rangle e^{2} } {T} \left [  \frac{4 \pi \rho_{ch}}{3} \right ]^{1/3}\ .
\label{eq:gammaMCP}
\end{equation}

Mixtures, having different charge densities, will generally not share a common $\Gamma_{crit}$ when considering the screened Coulomb repulsion in Eq. \ref{eq.V}. Corrections to melting temperature and $\Gamma_{crit}$ are calculated from a dimensionless screening parameter $\kappa = a / \lambda $ (the screening length and Wigner-Sietz radius) which for our mixtures fall between 1.7 and 2.4, of the same order as previous MD simulations studying astrophysical Coulomb crystals \cite{Hughto2011}. Corrections in the literature, such as Eqs. 17-19 in Hamaguchi and Eq 4. in Vaulina \etal\, are now widely cited and agree with each other to order $10^{-2}$ for $\kappa < 2 $ \cite{PhysRevE.66.016404,PhysRevE.56.4671}. In this work, we use the corrections provided by Eqs. 18 in Hamaguchi, which raise $\Gamma_{crit}$ from 175 to between 232 and 253 for the mixtures considered here. This is fairly small considering the range of $\Gamma_i$'s considered in a given mixture in this work (which span an order of magnitude), but is sufficiently large that it must be taken into account.

Observe that a MCP with a low $\Gamma_{crit}/\Gamma_{MCP}$ (\ie\ solid)  can have components of low charge, such that $\Gamma_{crit}/\Gamma_i >1$ (\ie\ liquid) is possible. It is these `liquid-like' ions present in mixtures, and their effect on the crystal structure of the MCP, that we seek to study. As low $Z_i$ nuclei have previously been identified as interstial defects by inspection in simulations by Horowitz \etal\, we are motivated to quantify the degree of `liquidness' of these light nuclei in this work.

\section{Simulations}\label{sec:sim}

\subsection{One-component Simulations}

\begin{figure*}[htp!]  %left bottom right top)
\centering
\begin{minipage}{.48\linewidth}
\includegraphics[width=0.99\textwidth]{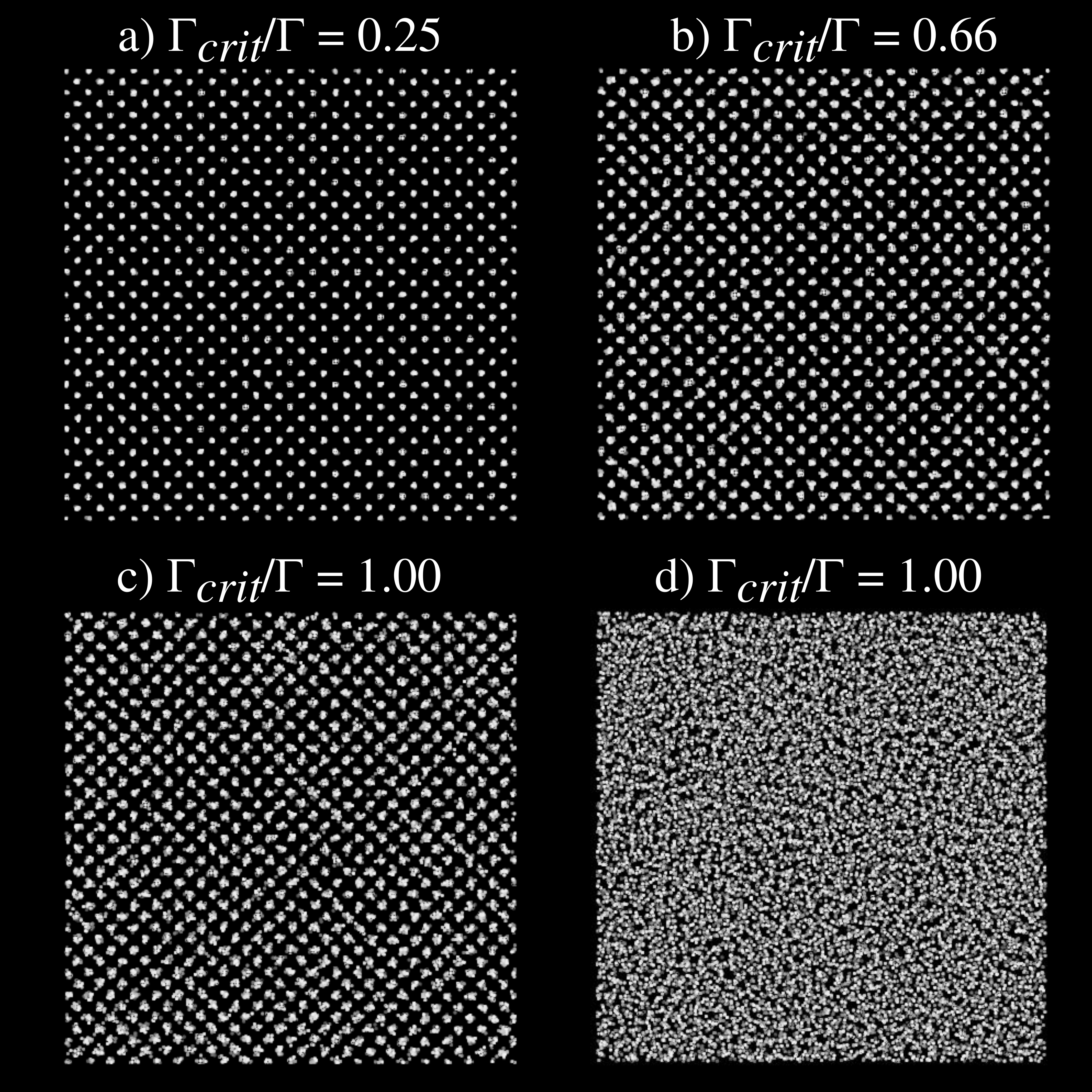}
\caption{\label{fig:OCP_MD} MD visualizations of an orthographic projection of the bcc (100) plane in an OCP. (a) At the lowest temperatures we find that the lattice has converged on a nearly ideal bcc structure. (b) For greater temperatures, we resolve the thermal fluctuations in the lattice resulting in a smearing of points. (c) At the melting temperature, the thermal fluctuations on lattice sites produce displacements comparable to the lattice spacing, while (d) the melted system at the same temperature demonstrates minimal order.}	
\end{minipage}\hfill
\begin{minipage}{.48\linewidth}
\includegraphics[trim=99 239 99 240,clip,width=0.99\textwidth]{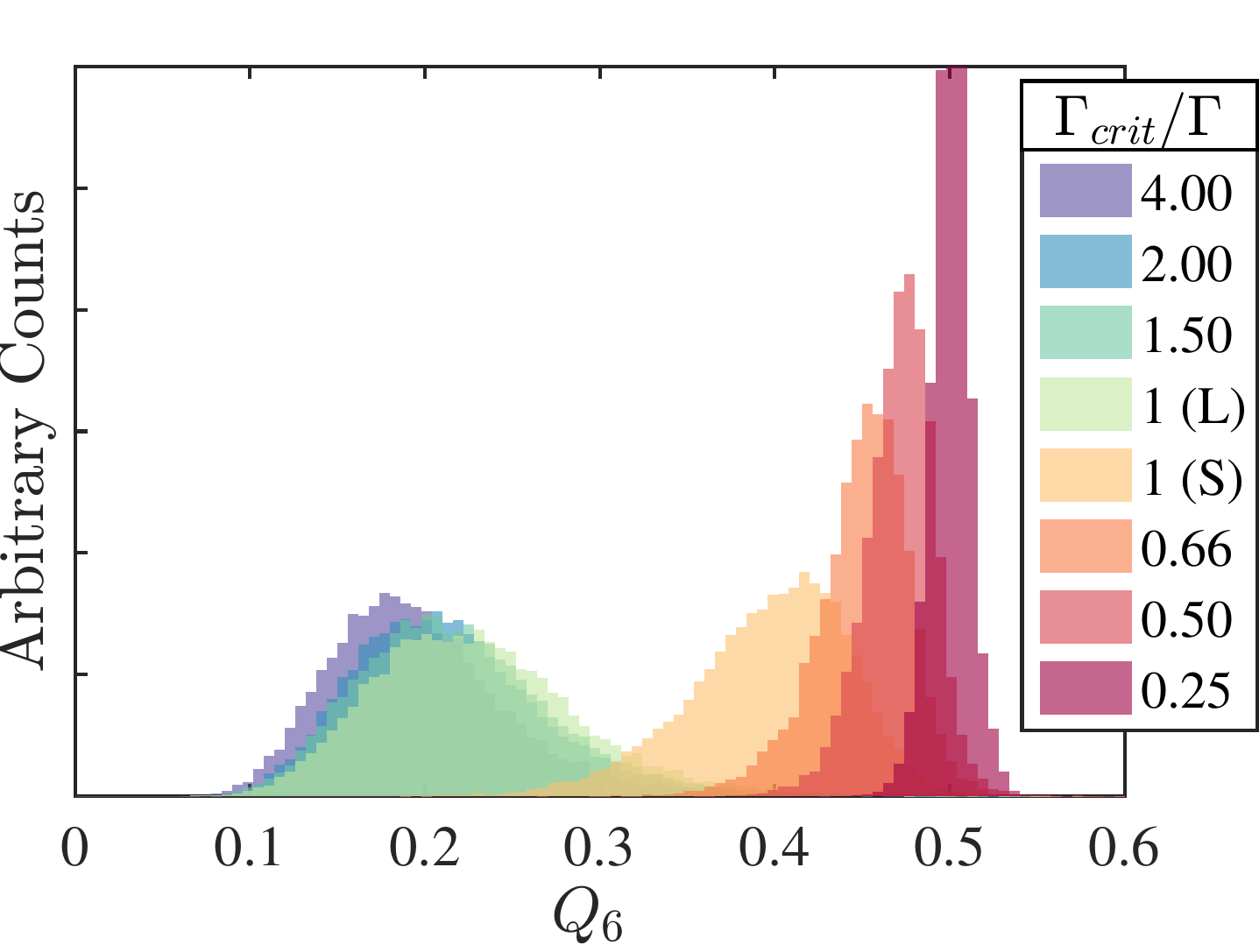}
\caption{\label{fig:OCP_Q6} (Color online) Histograms of ion bond order parameter $Q_6$ for a set of OCPs. While the liquids all demonstrate a mean $Q_6 \approx 0.2$, for solids the $Q_6$ varies with temperature between approximately 0.4 and 0.5, where the tightening of the distribution is understood as a reduction in thermal fluctuations on the lattice. The two simulations at $\Gamma_{crit}/ \Gamma =1$ are solid (S) and liquid (L) respectively and run at the melting temperature. A deficit of ions with $Q_6 \sim 0.3$ suggests that the bond order parameter is useful for discriminating between solid-like and liquid-like ions in pure phases based solely on the arrangement of their neighbors.} 
\end{minipage}\hfill
\begin{minipage}{.48\linewidth}
    \includegraphics[trim=0 20 20 20,clip,width=0.99\textwidth]{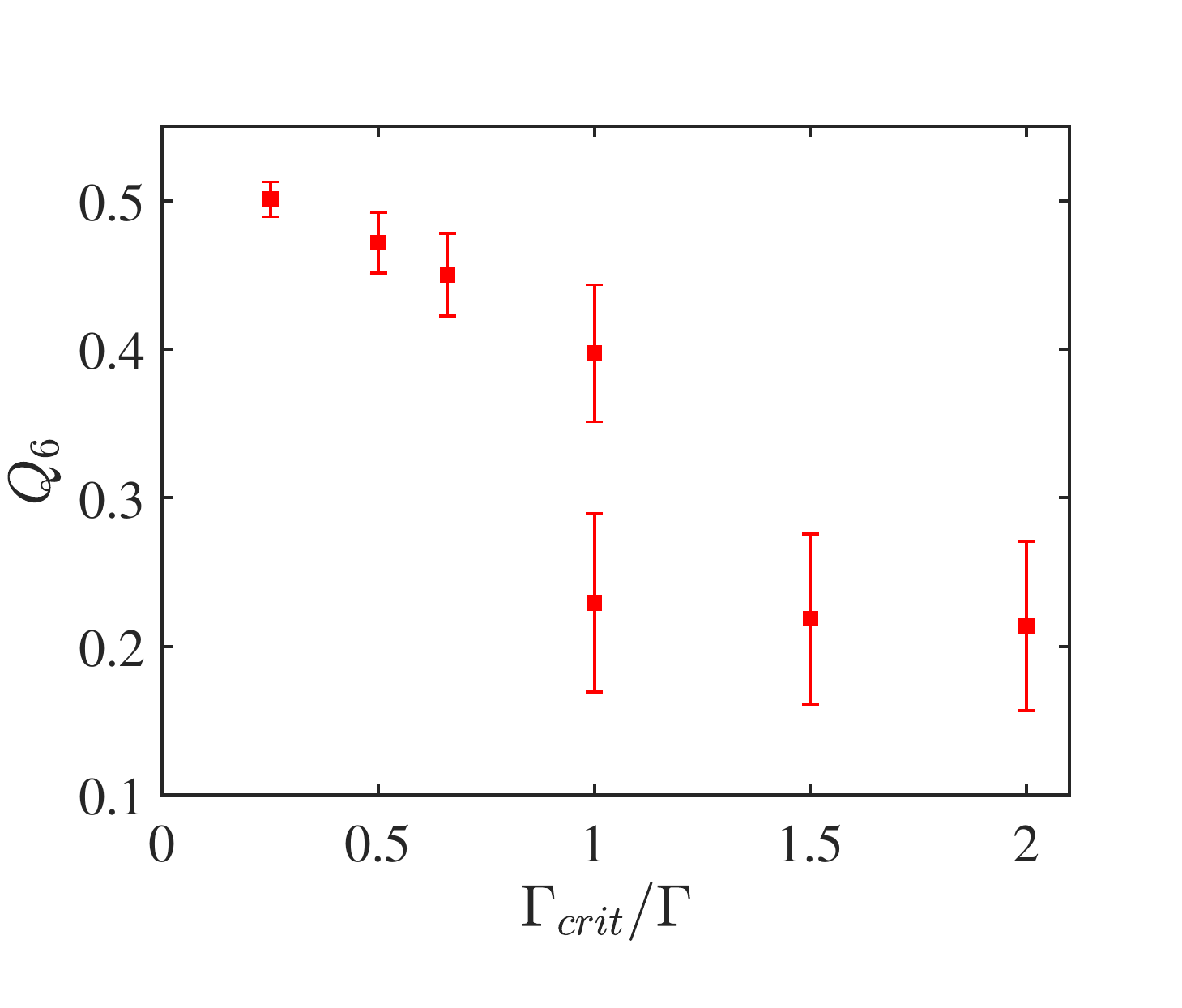}
    \caption{$Q_6$ for our OCPs. Points and error bars are mean and standard deviation of $Q_6$ distributions shown in Fig. \ref{fig:OCP_Q6}. We resolve both the sharpening of the distribution for solids with low $\Gamma_{crit}/\Gamma$ and the approximately constant behavior for liquids with high $\Gamma_{crit}/\Gamma$. Compare with the results for the MCP below.}
    \label{fig:OCP_Q6_scat}
\end{minipage}\hfill
\begin{minipage}{.48\linewidth}
    \includegraphics[trim=95 240 85 230,clip,width=0.99\textwidth]{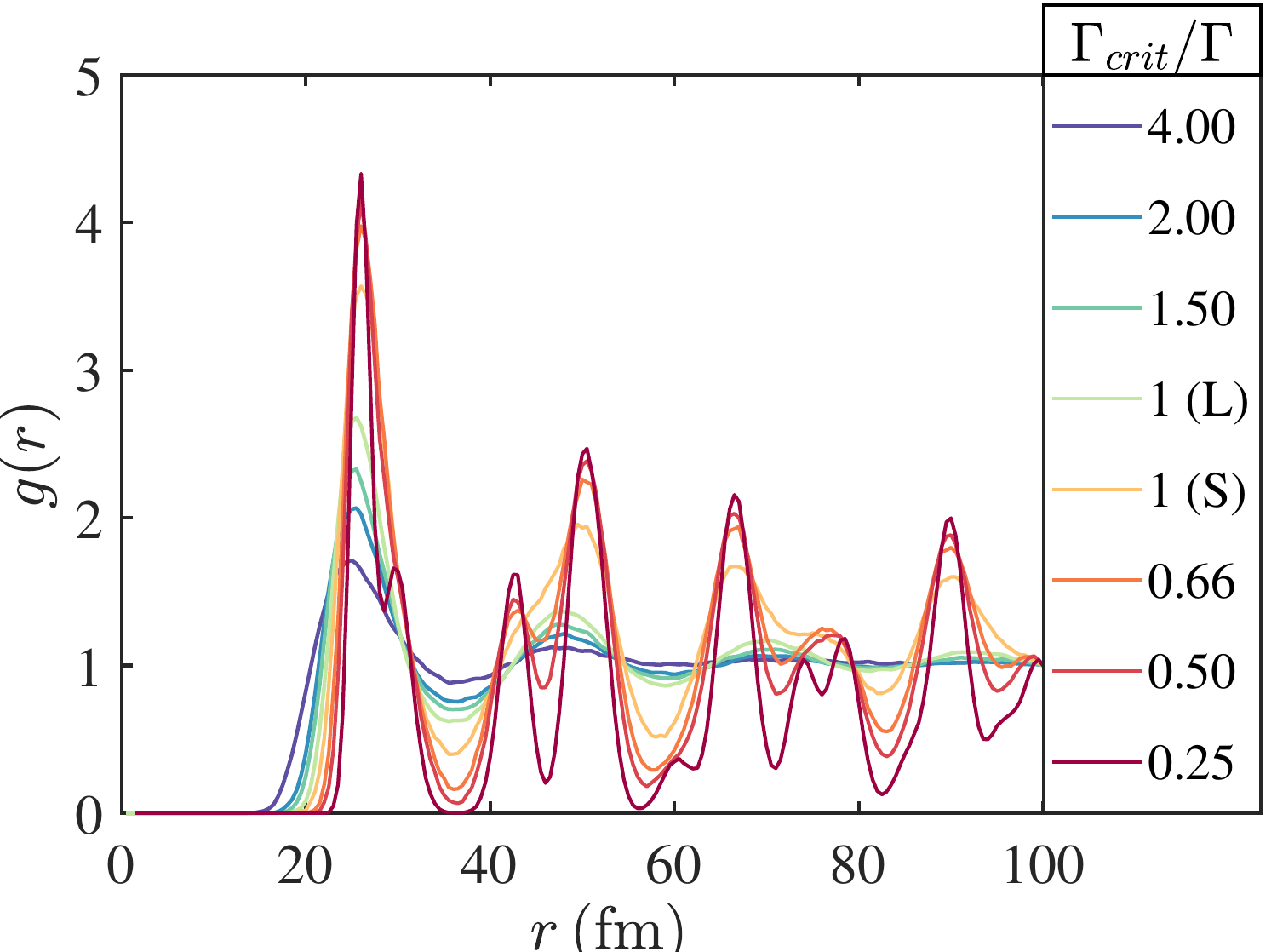}
    \caption{(Color online) Radial distribution functions $g(r)$ for the OCP. Color scheme and labeling is the same as Fig. \ref{fig:OCP_Q6}. The function broadens smoothly with increasing $\Gamma_{crit}/\Gamma$ between the known results for a bcc solid and liquid, with the transition at $\Gamma_{crit}/\Gamma = 1$ being resolved clearly. }
    \label{fig:grOCP}
\end{minipage}
\end{figure*}

%\begin{figure}[h!]
%    \centering

%\end{figure}

We perform simulations of 16,000 ions ($20 \times 20 \times 20$ bcc unit cells) in a cubic simulation volume with periodic boundary conditions at $ \Gamma_{crit} / \Gamma =$ 0.25, 0.50, 0.66, 0.97 1.00, 1.03, 1.33, 1.50, 2, and 4. Simulations $ \Gamma_{crit} / \Gamma > 1$ are liquid, while $ \Gamma_{crit} / \Gamma < 1$ are solid. Initial conditions for the liquid were taken to be random ion positions in the simulation volume, while the solid was taken to be a bcc lattice whose planes are aligned with the simulation boundaries. Visualizations are shown in Fig. \ref{fig:OCP_MD}.

For $ \Gamma_{crit} / \Gamma = 1$ we run two simulations, one using random initial conditions and one using lattice initial conditions. At the critical temperature both solid and liquid phases can exist allowing a useful temperature-independent comparison of solid-like and liquid-like behavior. All simulations are run at constant temperature for $10^6$ MD timesteps with temperature renormalizations every 100 timesteps following the procedure described in Caplan \etal \cite{Caplan2018}. All simulations are run using a timestep of 100 fm/c, and an ion density of $7.18 \times 10^{-5} \textrm{ fm}^{-3}$.\footnote{This ion density is largely irrelevant as $\Gamma$ is sufficient to fully describe the system, however it does determine the nearest neighbor spacing and gives the first minimum in the bcc radial distribution function at 35 fm.} The total energy converges within the first few thousand timesteps for all simulations, with fluctuations of order $10^{-5} E_{tot}$, indicating that configurations quickly reach thermal equilibrium. Total energy remained approximately constant with fluctuations of order $\Delta E/E \sim 10^{-6}$ during the equilibrium phase.

We calculate the bond order parameter $Q_6$ for each ion in the final configurations of these simulations, shown in Fig. \ref{fig:OCP_Q6} and Fig. \ref{fig:OCP_Q6_scat}. 
We calculate the local bond order parameter to ions within 35 fm, which corresponds to the first minima in the radial distribution function (Fig. \ref{fig:grOCP}) and thus contains the fourteen nearest neighbors (\ie\ each ion is at the center of a bcc unit cell of eight nuclei, with six nuclei at the center of each adjacent cell).

In Fig. \ref{fig:OCP_Q6} and \ref{fig:OCP_Q6_scat}, for $\Gamma_{crit}/ \Gamma \geq 1$, we observe that the bond order parameter $Q_6$ forms an approximately Gaussian distribution with a mean near $Q_6 \approx 0.2$. This is expected for a liquid with minimal local order. For $\Gamma_{crit}/ \Gamma \leq 1$, we observe a strong $\Gamma$ dependence in the distribution mean and width. For $\Gamma_{crit}/ \Gamma \lesssim 1$, the distribution shifts quickly toward $Q_6 \approx 0.4$ as $\Gamma_{crit}\Gamma$ decreases. As  $\Gamma_{crit}/ \Gamma$ approaches zero the distribution becomes strongly peaked around $Q_6 \approx 0.5$. Notably, a gap is observed near $Q_6 \approx 0.3$. This region of the histogram is largely unpopulated except for by the tails of a few distributions which are near the critical temperature. This strong separation in $Q_6$ for the OCP suggests that $Q_6$ will be a useful tool for discriminating between the solidness and liquidness of ions in MCPs. This technique for discriminating between solid and liquid ions has previously been used in studies of MCP phase separation by \citealt{Caplan2018}, though this method was not rigorously developed in that work. 
The radial distribution function $g(r)$ shows the known results for bcc solids at low $\Gamma_{crit}/ \Gamma$ and liquids at high $\Gamma_{crit}/ \Gamma$. The first order phase transition is apparent in our two simulations at $\Gamma_{crit}/ \Gamma = 1$ by observing the sharpening of the peaks and the emergence of the `double peaked' peaks as opposed to the approximately sinusoidal behavior for the liquid. It is worth noting that only for the lowest $\Gamma_{crit}/ \Gamma$ simulations is $g(35 \textrm{ fm}) \approx 0$, so our calculations of $Q_6$ using this cut-off includes only on average 14 nucleons for high $\Gamma_{crit}/ \Gamma$ simulations. Nevertheless, this is a small effect which we do not study further.

We can understand the behavior of $Q_6$ in terms of the temperature and position-space distribution of ions, shown in Fig. \ref{fig:OCP_MD}. For liquids, $\Gamma_{crit}  / \Gamma \geq 1$, and we expect similar distributions of ions for all temperatures. A slight leftward skew (toward lower $Q_6$) that appears with increasing effective temperature (\ie\ greater $\Gamma_{crit}  / \Gamma$) may be interpreted as the effect of greater average thermal fluctuations. Meanwhile, for solids with $\Gamma_{crit}/ \Gamma \leq 1$, the reduction in thermal energy suppresses thermal fluctuations on lattice sites, resulting in a sharpening of the distribution of $Q_6$ as ions converge on an idealized lattice with high local order, as seen in Fig. \ref{fig:OCP_MD}.  

With structural characterizations for the OCP complete we move on to study the MCP.

%see thesis text for initialization, freezing, and 'equilibration/relaxation'

\subsection{Multicomponent Plasmas}\label{ss:mcp}

\subsubsection{Mixtures}\label{ss:mcp_mixtures}

\setlength{\tabcolsep}{0.8em} % for the horizontal padding
{\renewcommand{\arraystretch}{1.3}% for the vertical padding
\begin{table*}[th]\label{tab:mix}\caption{Summary of Mixtures}
\begin{tabular}{c c c c c c c c c c}
\hline  \hline  
\rule{0pt}{3ex} \vspace{0.1cm}     \hspace{2.8cm} & $\langle Z \rangle $ & $Q_{imp}$ & $\langle Z^{5/3} \rangle $ & $\Gamma_e$ & $\Gamma_{crit} $ & $Z_{max} (\Gamma_{crit} / \Gamma_{i})$ & $Z_{min} (\Gamma_{crit} / \Gamma_{i})$ & $\Gamma_{MCP}$ \\  \cline{2-10}  
Mixture no. 1 & 11.2  & 3.1       & 57.2        & 5.841     &  253.4  & 24  (0.21)         & 8  (1.36)           & 333.9          \\
Mixture no. 2 & 23.7  & 7.5       & 196.6       & 1.539     &  238.8  & 26  (0.68)         & 8  (4.85)           & 302.5          \\
Mixture no. 3 & 24.8  & 4.6       & 212.5       & 1.340     &  237.9  & 28  (0.69)         & 8  (5.55)           & 284.8          \\
Mixture no. 4 & 26.8  & 8.7       & 241.1       & 1.202     &  236.5  & 30  (0.68)         & 8  (6.14)           & 289.9          \\
Mixture no. 5 & 28.0  & 8.5       & 261.7       & 1.087     &  235.5  & 32  (0.67)         & 11 (3.98)           & 284.4          \\
Mixture no. 6 & 32.1  & 27.7      & 334.4       & 1.119     &  232.9  & 44  (0.38)         & 16 (2.05)           & 374.1\\     \hline   \hline   
\multicolumn{10}{p{.9\textwidth}}{ \vspace{0.05cm}    Full list of species and abundances for each mixture are available in the appendix. We report here: mean ion charge $\langle Z \rangle $, variance in ion charge $Q_{imp}$, mean of ion charge to the 5/3 power $\langle Z^{5/3} \rangle$, reference $\Gamma_e$ such that $\Gamma_i = Z_i^{5/3} \Gamma_e$ (which depends on temperature), the melting criteria $\Gamma_{crit}$, the greatest and smallest charges included in the mixture $Z_{max}$ and $Z_{min}$ and associated $\Gamma_{crit} / \Gamma_i$ for those species. Total mixture $\Gamma_{MCP} = \langle Z^{5/3} \rangle \Gamma_e$ is effectively the chosen temperature we simulate at.}
\end{tabular}
%\linebreak \vspace{0.05cm}\linebreak
\end{table*}
}

We study six mixtures from Mckinven \etal\ (2016) \cite{Mckinven2016}. That work calculated the phase separation that occurs for rp-ash mixtures in equilibrium that were 50\% solid and 50\% liquid. In this work we perform molecular dynamics simulations of only the solid component of those mixtures. This differs from past work which was concerned with simulating the phase separation which included both the solid and liquid components (see \citealt{Caplan2018}).  

These mixtures correspond to the burning products for six different accretion rates of solar composition (helium mass fraction $Y=0.2752$) material. These accretion rates are  $\dot{m}  / \dot{m}_{Edd}$ = 0.1, 0.2, 0.3, 0.5, 1.0, and 10 in units of the Eddington accretion rate $\dot{m}_{Edd}$. We refer to these mixtures hereafter as mixtures nos. 1-6 respectively, and refer to the concentrations (number abundances) of their components with charge $Z_i$ as $c_{Z_i}$. We note that the mixture no. 2  that we use in this work is different from what is reported in Mckinven \etal\ (\ie\ $\dot{m}_{Edd}=0.2$ in their Tab. 1). That work reports a high impurity parameter for the solid produced by phase separation of the parent mixture, with a large concentration of $Z=12$ ions in the solid. Further work on this mixture suggests that it is near a eutectic point, similar to the discussion in \citealt{Caplan2018}. We have recalculated the phase separation and find this mixture is depleted in $Z=12$ nuclei, producing a much purer solid which is comparable to the other mixtures reported in that work \cite{MckinvenComp2019}.

The phase separation of the parent mixtures that produce these solids are described in some detail in Tab. 1 and Fig. 1 of Mckinven \etal, and we include in the appendix a detailed list of the ions and abundances used in our calculations. We exclude species with concentrations less than $10^{-5}$. We treat all nuclei of the same charge as having the same mass, chosen as either the mass of the most abundant isotope, or as the average mass of all isotopes rounded to the nearest integer when several isotopes have comparable abundances (\ie\ we use $(Z,N+1)$ when pairs of even-even isotopes at $(Z,N)$ and $(Z,N+2)$ have comparable abundance). 

In Tab. \ref{tab:mix} we summarize some key features of the mixtures. It is worth noting that the $\Gamma_{MCP}$ reported here is different than $\Gamma_s$ in Tab. 1 of Mckinven, though one might naively expect these to be the same. The $\Gamma_s$ reported in that work is specifically the plasma parameter for the solid part of a 50-50 solid-liquid system that is in equilibrium following phase separation of the parent mixture. This is not necessarily the $\Gamma$ (\ie\ temperature) the lone solid mixture freezes at, and so our mixture here which excludes the liquid need not be simulated at this specific temperature.

%These simulations are all run for $4 \times 10^7$ timesteps with a timestep of 25 fm/c, at a density of $7.18 \times 10^{-5} \textit{ fm}^{-3}$ (as in \citealt{Horowitz2009}. Temperature is renormalized every 10 fm/c, as above. Temperatures are chosen to be  

\subsubsection{Simulations}

The preparation of these configurations is considerably more detailed than the OCP, as we want to study a realistic crystal which is not heavily biased by initial conditions. To briefly summarize, the simulations are initialized as a liquid with a uniform random distribution and cooled until they freeze to form a crystalline solid. This solid is then equilibrated (\ie\ allowed to evolve to equilibrium) at constant temperature. The effective temperature we simulate at, $\Gamma_{MCP}$, is between approximately $0.6\Gamma_{crit}$ and $0.8\Gamma_{crit}$.
%just below the melting temperature predicted by Mckinven \etal\. 

These six simulations all contain 102,400 ions at a density of 7.18 $\times 10^{-5} \textrm{ fm}^{-3}$, with a timestep of 25 fm/c, in a cubic volume with periodic boundaries as before. The one exception is the simulation of mixture no. 1, which includes 204,800 nucleons which also serves as a comparison to check for finite size effects. These simulations are all initialized from random positions and velocities are randomly generated with a Maxwell Boltzmann distribution whose temperature is chosen to be above the temperature given by Mckinven \etal, given as $\Gamma_s$ in that work. This produces liquid configurations which are simulated at constant temperature for at least $10^{6}$ MD timesteps ($2.5 \times 10^{7}$ fm/c). The simulations are then cooled by rescaling the velocities every 1000 timesteps to a Maxwell-Boltmann distribution to decrease the temperature by $5 \times 10^{-6}$ MeV. This cooling is simulated in intervals of $10^{6}$ timesteps, which continues until the configuration freezes. The instant of freezing is straightforward to identify as the energy per particle shows a sharp decline consistent with a first order phase transition and the lattice structure becomes visible by inspection, as in Fig, \ref{fig:OCP_MD}. This general equilibration scheme has been used extensively in past work \cite{Horowitz2007,Horowitz2009,Caplan2018}.

The simulation configurations generated immediately after the phase transition (\ie\ the highest temperature we are certain the solid is stable and will not spontaneously melt) is then evolved at constant temperature for $8 \times 10^{8}$ timesteps ($2 \times 10^{9}$ fm/c) over two simulations of $4 \times 10^{8}$ timesteps ($10^{9}$ fm/c) each. Over the first simulation the energy is observed to asymptotically decrease, suggesting that the newly formed solid is relaxing to an equilibrium. Over the second simulation we observe that the total energy is constant, suggesting that our six configurations have equilibrated. 

As the static structure factor $S(q)$ has been used extensively in past work it is appropriate to comment here on our choice to not report $S(q)$ here. This work is considering high lattice temperatures (\ie\ $T \lesssim T_{melt}$) which are significantly greater than the lattice Debye temperature. This implies that the thermal conductivity is dominated by electron-phonon scattering in the astrophysically relevant regime, as in \citealt{Deibel_2015} (see also \citealt{Potekhin1999}). Any static structure factors reported in this work would be phonon dominated, rather than impurity dominated. Future work may be interested in studying the $S(q)$ for the configurations generated in this work quenched to low temperature.

%Deibel_2015 

\subsubsection{Crystal Structure}

Ions in our simulations crystallize and form a bcc lattice, as in the OCP case. From inspection, the crystals formed in our simulations of mixtures nos. 2, 3, and 5 appear without any immediately apparent structural defects such as dislocations or grain boundaries, while mixtures nos. 1, 4, 6 may contain dislocations, but are otherwise perfect bcc crystals. %The interpretation of crystalline in defects in mixtures no. 6 is supported by calculations of the principal shear stresses which find that $abs{sigma_{ij}} \sim 10^-6 \text{MeV/fm}^-3$ for all three principal shear stresses which suggest that some small disorder is frozen in.  %%PROB WRONG, MIX 2 ALSO HAS ANISOTROPY IN PRINCIPAL STRESSES
This is expected to have only a small effect on the bond order parameters we calculate, as a planar defect in a cubic volume will only involve order $N^{2/3}$ ions; for simulations with 102,400 ions this effect is of order $10^{-2}$.

\begin{figure}[t!]
\centering
\includegraphics[width=0.47\textwidth]{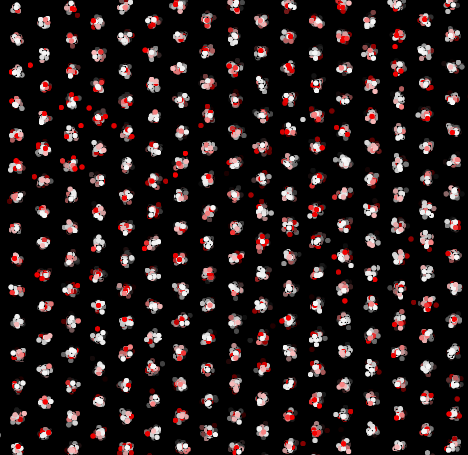}
\caption{\label{fig:MD} (Color online) Subvolume of a molecular dynamics simulation of mixture no. 3. Each point represents one ion. The simulation is fully 3D, though we present an orthographic projection of the bcc (100) face here for clarity. The clustering on lattice sites is due to the distribution of ions in the third dimension. The crystal forms a bcc lattice with a single domain and without any immediately apparent structural defects. Ions with charge near the mixture average and above are shown in white, while those with low charge appear in red. The colors are dimmed with increasing depth in the field, and the red-saturation shows the relative charge. Note that most interstitial ions are red, particularly in top left.}	
\end{figure}

\begin{figure*}[htp!]
\centering  %left bottom right top)
\includegraphics[trim=75 40 65 20,clip,width=0.98\textwidth]{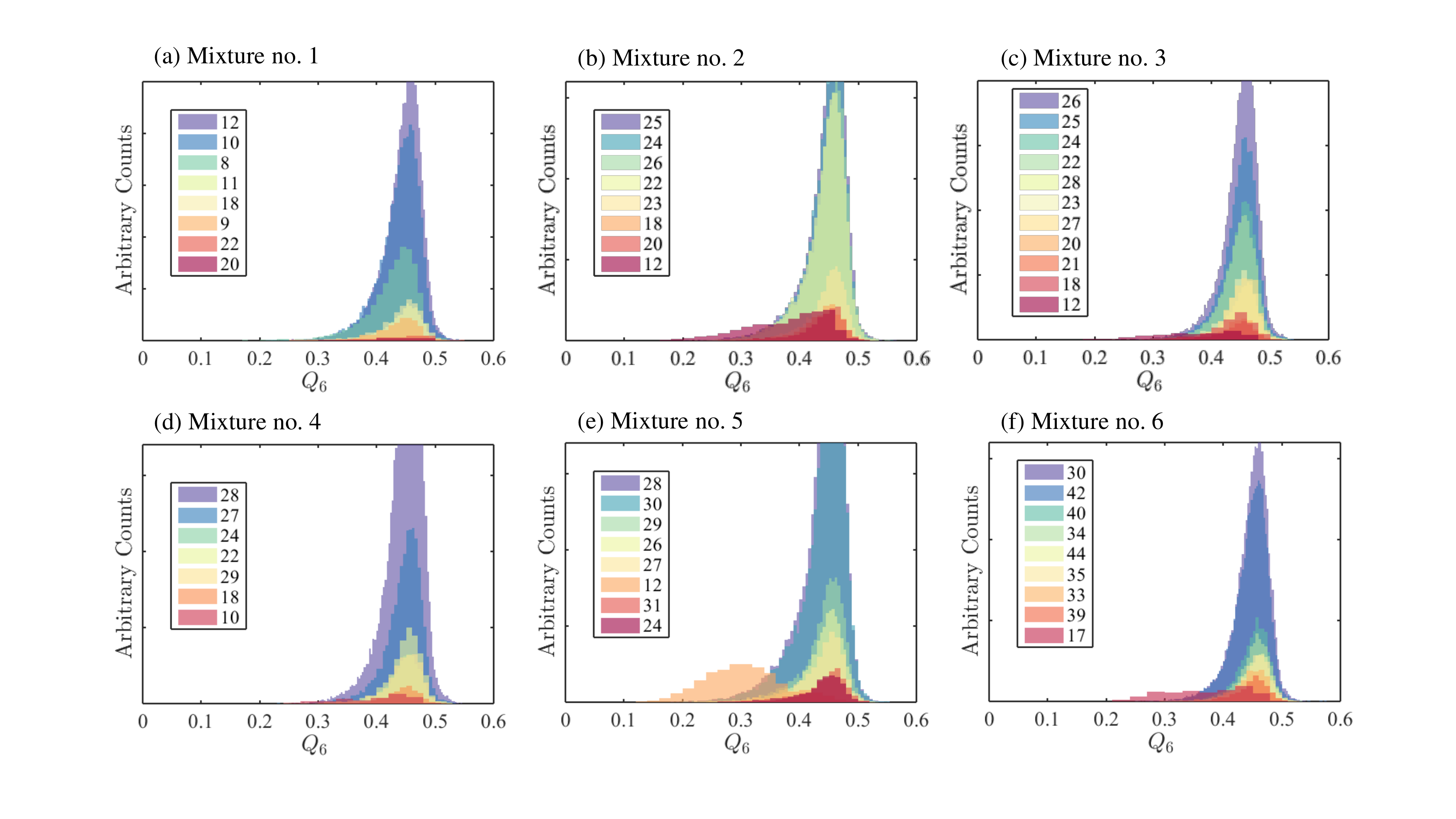}
\caption{\label{fig:6MCP_BOP} (Color online) Histograms of ion bond order parameter $Q_6$ for select ion species in our six MCPs. Legends show ion charge $Z$, in decreasing order of abundance. Relative heights of distributions show the relative abundance of ions within a mixture, though some bin widths for the least abundant species have been rescaled for readability. Compared to the OCP, we find a greater left skew in our distributions. Furthermore, the lowest $Z$ species (\ie\ those with $\Gamma_{crit}/\Gamma_i > 1 )$ in the mixture have much broader distributions with lower average $Q_6$ relative to the mixture average.} \vspace{7mm}
\begin{minipage}{.48\linewidth}
\includegraphics[trim=60 216 85 230,clip,width=0.99\textwidth]{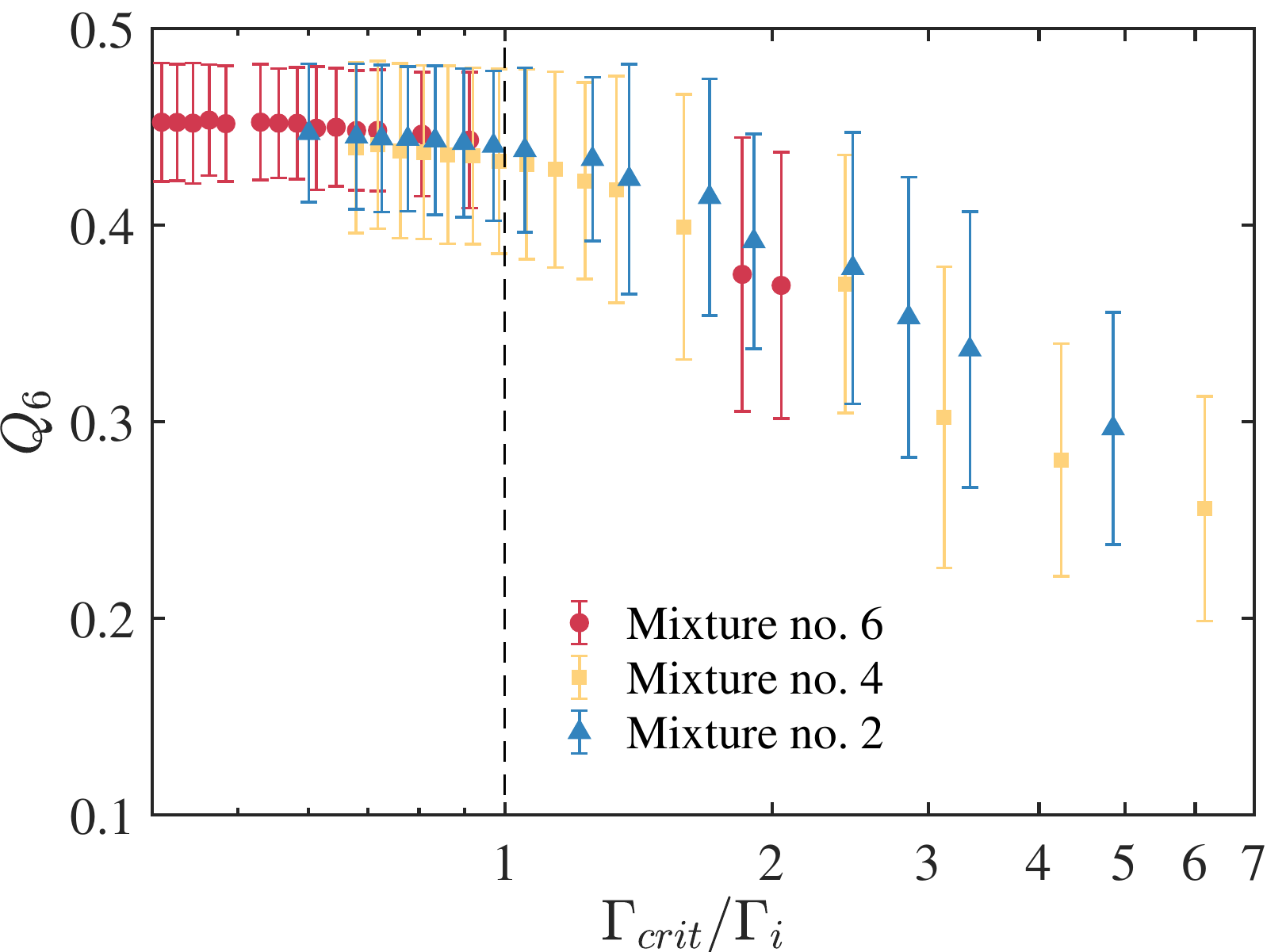}
\caption{\label{fig:mix4} (Color online) $Q_6$ for all species in mixture nos. 2, 4, and 6 (nos. 1, 3, and 5 omitted for readability). Points and error bars are the average and standard deviation of $Q_6$ for ions of that respective species. Species with charges below the critical value for the mixture (vertical dashed line) all show similar solid-like behavior as $\Gamma_{crit}/\Gamma > 1$, as in Fig. \ref{fig:6MCP_BOP}, while ions with $\Gamma_{crit}/\Gamma < 1$ are increasingly liquid-like $Q_6$. Compare with Fig. \ref{fig:OCP_Q6_scat}.}
\end{minipage}\hfill
\begin{minipage}{.48\linewidth}
\includegraphics[trim=15 25 -5 30,clip,width=0.99\textwidth]{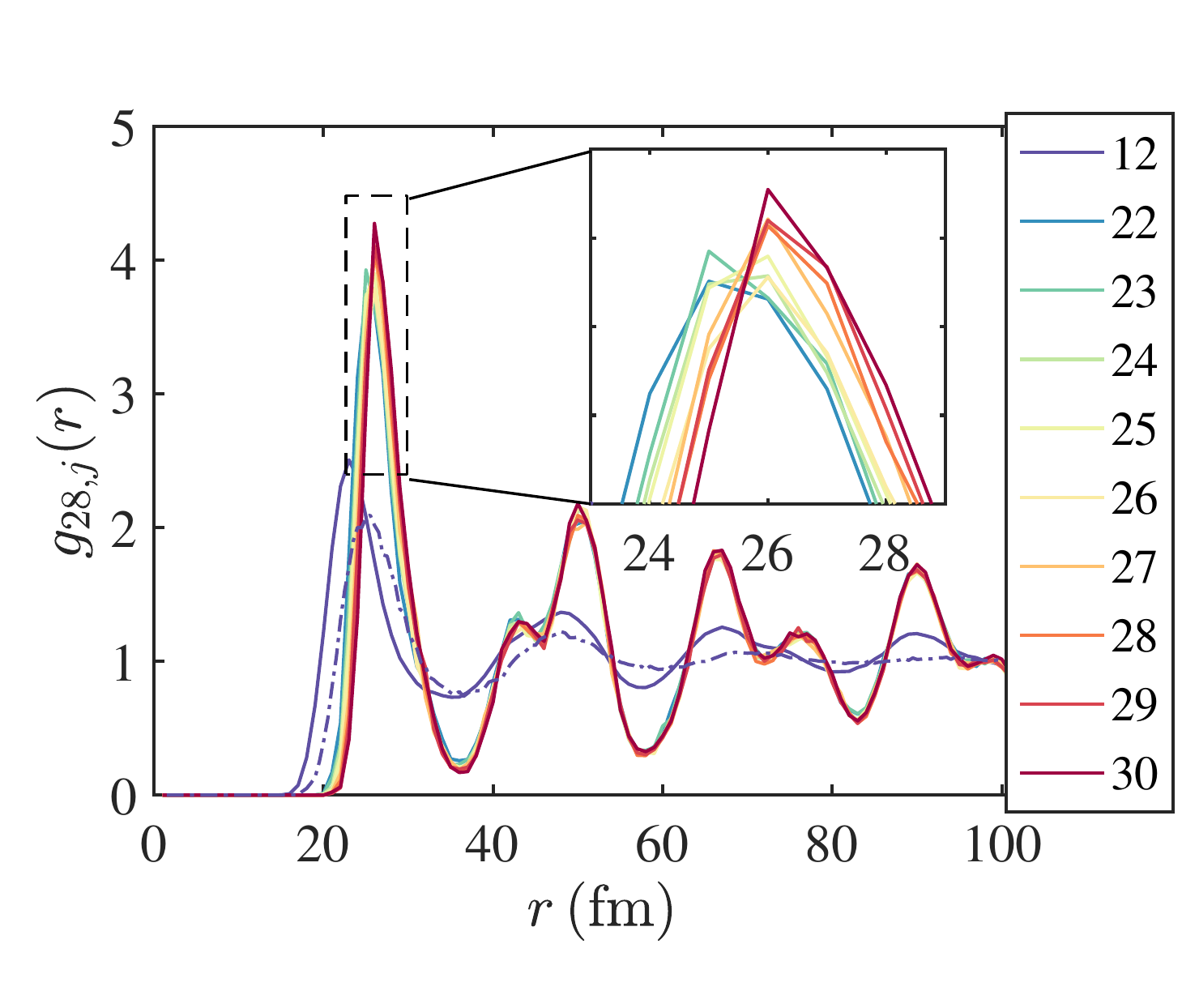}
\caption{\label{fig:gr} (Color online) Radial distribution functions $g(r)$ for mixture no. 4 for all species $j$ relative to the dominant mixture component, $Z=28$. Species are shown in order of increasing charge for ease of comparison with the OCP in Fig. \ref{fig:grOCP} ($\Gamma_{crit}/\Gamma\propto Z_i^{-1}$). Observe that the lowest charge plotted ($Z=12$,  $\Gamma_{crit}/\Gamma_i = 3.2$, dark purple) shows liquid characteristics, compare to the OCP at $\Gamma_{crit}/\Gamma = 1.5$ (dashed purple). All other species are consistent the known result for a bcc solid at  $\Gamma_{crit}/\Gamma_{MCP}$. Inset shows the separation in the first peak with charge, discussed in the text.}	
\end{minipage}
\end{figure*}

We find little evidence for continued phase separation which suggests that these mixtures may be stable under the conditions we simulate. We show a subvolume of one configuration generated during the simulation of mixture no. 3 in Fig. \ref{fig:MD}. The most abundant ion species in this mixture has $Z=26$ with concentration $c_{26} = 0.427$. Ions with charges of $Z=26$ or greater are shown in white, while ions with decreasing charge are shown with increasing redness. Ions with charge $Z \leq 22 $ have a total concentration of 0.152 and are shown in red. Most of these red points do appear on lattice sites indicating that low $Z$ impurities are not necessarily interstitial, nevertheless the few interstitial points that are easily identifiable are all low $Z$ ions. In the upper left region of the figure there is a cluster of interstitial low $Z$ ions. This is largely due to the projection in the third dimension; many of these points are well separated, though it could be taken to be evidence of clustering of light nuclei and further phase separation, as seen by \citealt{Horowitz2009}.

We show $Q_6$ for select species in these mixtures in Fig. \ref{fig:6MCP_BOP}. We have separated $Q_6$ by species in each mixture to show the trend with species charge $Z$ (identified in the legend). The legends present species in order of decreasing abundances, so that approximate abundances can be seen in the relative heights of the histograms (most abundant in purple, least abundant in red). For readability, some low abundance species have their bin widths rescaled and we omit a number of species from these plots. We choose to show only the most abundant species as well as those which clearly demonstrate the general behavior of low $Z$ species. 

Direct comparisons between mixtures may be difficult as mixtures have different average charge and were simulated at different temperatures. Nevertheless, trends are apparent. In every mixture high $Z$ ions have a high average $Q_6 (\approx 0.45)$, as in the solid OCP, but these distributions now have left skew. Ions with $Z$ much lower than the mixture average (most visibly $Z=12$ in mixture no. 3 and no. 5) show broader distributions with lower average $Q_6 (\approx 0.3)$. Physically, high $Z$ ions all have more regularly arranged nearest neighbors. Below some threshold in $Z$, ion nearest neighbors become less regular, and low $Z$ ions are found at centers of local disorder.

%low $Z$ ions show broader distributions shifted to lower $Q_6$

%In a given mixture the local bond order parameter for nuclei of the same species have an approximately Gaussian distribution with left skew. Error bars show $1 \sigma$ standard deviation.

%High $Z$ ions demonstrate high Q_6%

%(Nearest neighbors are taken to be ions within 35 fm)
% ./bop 35 0 < md.02056250000.xyz > lbop.dbg

In Fig. \ref{fig:mix4} we present more detailed information about $Q_6$ for mixture nos. 2 (blue triangles), 4 (light yellow squares), and 6 (dark red circles). These are representative of all six mixtures (mixtures 1, 3, and 5 are omitted for readability) and a log-scale is used to avoid overcrowding the points for $\Gamma_{crit}/\Gamma_i < 1$. We plot all components of these three mixtures, resolving more clearly the intermediate behavior of trace low $Z$ (high $\Gamma_{crit}/\Gamma_i$) species. 
%To compare with the OCP in Fig. \ref{fig:OCP_Q6_scat}, recall that increasing $Z_i$ corresponds to an effectively decreased $\Gamma_{crit} / \Gamma_i$, so that if the MCP were just a linear sum of its component OCPs we would expect the average $\bar{Q_6}$ to increase asymptotically with $Z$ to $\bar{Q_6} \approx 0.5$ as $\Gamma_{crit}/\Gamma_i \rightarrow 0$, while the width of its distribution $\sigma_{Q_6}$ should decrease. %Consistent with the OCP, we find that $\bar{Q_6}$ is largely independent of charge for large $Z$, with $\bar{Q_6}\approx 0.45$. While this value is consistent the result for an OCP near the melting temperature shown above, it is unlike the OCP which shows a sharpening in that distribution; $\sigma_{Q_6}$ here is approximately constant. 
In contrast to the OCP, all species for which $\Gamma_{crit} / \Gamma_i < 1$ show approximately the same behavior, which is approximately that of the mixture average, \ie\ there is some equivalent $\Gamma_{OCP}$ which describes high $Z$ ions in this mixture such that  $\Gamma_{OCP} \approx \Gamma_{MCP}$. If the MCP were just a linear sum of its component OCPs we would expect the average $\bar{Q_6}$ to increase asymptotically with $Z$ to $\bar{Q_6} \approx 0.5$ as $\Gamma_{crit}/\Gamma_i \rightarrow 0$, while the width of its distribution $\sigma_{Q_6}$ should decrease.
Furthermore, near  $\Gamma_{crit} / \Gamma_i = 1$, $\bar{Q_6}$ begins smoothly decreasing. This intermediate behavior may be analogous to a glass transition, or indicative of a similar second order phase transition.

In Fig. \ref{fig:gr} we show the radial distribution functions $g_{i,j}(r)$ for mixture no. 4, again taken to be representative of all our mixtures. These $g_{i,j}(r)$ are calculated between all ions of species $j$ and all ions of species $i$, where species $i$ is $Z=28$ which comprises more than half of the mixture. Qualitatively, these are the pairwise distributions between the dominant species of the mixture and all other species. Species abundant to less than $c_j < 4 \times 10^{-3}$ are excluded due to poor statistics. Observe that $g_{28,12}(r)$ (solid dark purple), for the lowest charge species in the mixture $Z=12$ ($\Gamma_{crit}/\Gamma_i = 3.2)$, is the known result for a liquid. We include a similar OCP, with $\Gamma_{crit}/\Gamma=1.5$ (dashed dark purple) for comparison. We observe that despite having a higher effective temperature, $g_{28,12}(r)$ shows stronger correlations with higher order nearest neighbors than the true liquid as they are embedded in a solid lattice. All other species shown here demonstrate the known result for a bcc solid at finite temperature. This supports our interpretation of Fig. \ref{fig:6MCP_BOP} above which argues that low $Z$ species are in a liquid-like state within the lattice. In the inset, the first peak of $g(r)$ shows a separation in charge with the first peak shifting to greater $r$ for greater $Z$. However, no such separation is observed for higher order peaks. We argue that this is evidence of screening on the lattice. Lattice spacing is preserved out to large $r$, so the lattice remains ordered over a large number of lattice sites. However, the shifting in nearest neighbors may be due to the Coulomb repulsion between individual ions. Lower $Z$ ions have weaker Coulomb repulsion relative to the lattice, and so they may be closer to high charge neighbors than the average lattice spacing, while high $Z$ ions have greater Coulomb repulsion and thus will have greater average separation from neighbors. The absence of any separation in the second order peaks or higher suggests that this is a local effect where high $Z$ ions are screened by low $Z$ nearest neighbors so that neighboring cells have an average charge close to the mixture average.

\section{\label{sec:dis} Discussion}

We characterize the structure of the OCP and MCP near the melting temperature using the bond order parameter $Q_6$ and radial distribution function $g(r)$, finding generally that the structure of the MCP is more complicated than a linear sum of the OCP behavior of its components. We interpret the behavior of the majority of species in our mixtures to be that of a solid having crystal properties similar to an OCP with equivalent $\Gamma_{OCP} = \Gamma_{MCP}$, \ie\ the bond order parameter and radial distribution functions are effectively those of the one component plasma at the mixture average temperature for species where $\Gamma_{crit} / \Gamma_i < 1$. Trace low $Z$ species for which $\Gamma_{crit} / \Gamma_i > 1$ show intermediate behavior between what was shown for solid and liquid OCPs, including liquid-like radial distributions relative to the lattice average and low average values for the bond order parameter. This `quasi-liquid' behavior suggests that low $Z$ ions may congregate together in regions with low local order, such as near grain boundaries, dislocations, or in local pockets of liquid embedded within the solid. These sorts of structural defects have been reported for the MCPs simulated by Hughto \etal\ and Caplan \etal \cite{Hughto2011,Caplan2018}. 

%Screening:
We observe screening effects in our simulations of MCPs, with nearest neighbor separations being affected by ion charge but higher order neighbors all being found at the lattice average separation regardless of charge. This screening behavior may be relevant to calculations of the transport properties of the crystal, in particular the thermal and electrical conductivities. Many astrophysical models rely heavily on the impurity parameter $Q_{imp} = (1 / n) \Sigma_i n_i ( \bar{Z} - Z_i)^2 $ defined as the variance in mixture charge, which does not contain information about the lattice structure \cite{deBlasio1998}. The impurity parameter formalism was developed assuming a small number of impurities randomly distributed in a relatively pure lattice and may not generalize to mixture with many components of similar abundance. Recent work by \citealt{Roggero2016} finds that, for mixtures similar to those studied in this work, the effective impurity parameter when accounting for lattice effects is a factor of 2-4 lower than would be predicted from $Q_{imp}$ alone. We explain this physically in terms of the lattice structure. Low $Z$ and high $Z$ `impurities' may tend to fall on adjacent lattice sites, screening each other and preserving long range order in the lattice. For accurate calculations of the transport properties of MCPs past work has relied on computationally expensive MD and PIMC simulations. Taken together with the work by Roggero, these calculations may motivate theoretical work to efficiently determine effective transport properties for a given mixture knowing only the composition which do not rely on computationally expensive simulations. 

Future work may seek to study how the lattice formed by these mixtures evolves when annealed to lower temperatures. For example, as $\Gamma_{crit} / \Gamma_i \ll 1 $ for all species, the low $Z$ interstitial defects may either be frozen in or they may migrate to lattice sites, though this will be difficult to study directly with molecular dynamics simulations owing to the long equilibration times and low diffusion rates at low temperatures. Still, such simulations may be interesting and their static structure factors may provide useful insight for improving estimates of the effective impurity parameter in accreting neutron stars.

%Diffusion:
This work may motivate future studies of structural properties in MCPs, such as diffusion, which is relatively unstudied in the literature. Though this is conjecture, the intermediate behavior in $Q_6$ observed for low $Z$ species may generalize to other crystalline properties, such as diffusion coefficients. Following from linear mixing theory, low $Z$ ions likely have higher mobility within the lattice, having less Coulomb energy relative to the lattice average. Their larger (relative) thermal energies may raise their diffusion coefficients relative to the lattice average, having a greater tunneling probability for lattice site hops. For example, Hughto \etal\ calculated diffusion coefficients for the OCP and found that the diffusion coefficients for a solid at the melting temperature are two orders of magnitude lower than for a liquid \cite{Hughto2011}. We conjecture that an intermediate diffusion regime exists in MCPs near the melting temperature which may be studied in future work, where low $Z$ constituents of the solid with $\Gamma_{crit}/\Gamma_i > 1$ diffuse almost freely within the lattice, but with suppressed diffusion relative to an purely liquid OCP due to the rigid lattice structure. If diffusion rates evolve smoothly near $\Gamma_{crit}/\Gamma_i = 1$ in mixtures, then there could be a number of implications for astrophysics. Such diffusion rates would be relevant for the evolution of the crystal structure and composition both during freezing and as the mixture evolves to lower effective temperature (\ie\ lower $\Gamma_{crit}/\Gamma_{MCP}$), specifically in freezing white dwarfs and accreting neutron stars.

\textit{Acknowledgements}: We thank C. Horowitz, A. Cumming, R. Mckinven, and E. Brown for helpful conversations and the Canadian Institute for Theoretical Astrophysics and the Institute of Nuclear Theory for hospitality. This work benefited from support by the National Science Foundation under Grant No.  PHY-1430152 (JINA Center for the Evolution of the Elements). The authors acknowledge the Indiana University Pervasive Technology Institute for providing HPC (Big Red II) and storage resources that have contributed to the research results reported within this paper. This research was supported in part by Lilly Endowment, Inc., through its support for the Indiana University Pervasive Technology Institute, and in part by the Indiana METACyt Initiative. The Indiana METACyt Initiative at IU was also supported in part by Lilly Endowment, Inc. This material is based upon work supported by the National Science Foundation under Grant No.  CNS-0521433.

%\nocite{*}
\bibliography{aipsamp}% Produces the bibliography via BibTeX.

%\newpage

\appendix 

\section{Mixture Tables}

\setlength{\tabcolsep}{0.8em} % for the horizontal padding
{\renewcommand{\arraystretch}{1.3}% for the vertical padding
\begin{table*}[]
\caption{\label{tab:appendix} Complete list of mixtures studied in this work, with ion charge $Z$, ion mass $A$, and total number abundance $N$ and fractional number abundance $n = N/102400$. As described in the text, we take all ions of the same charge to have the same mass, though real mixtures will have a range of isotopes.}
%\hline \hline
\begin{tabular}{cc|cc}
\multicolumn{4}{ c}{Mixture no.  1} \\
\hline \hline
$Z$  & $A$  & $N$     & $n$       \\ \cline{1-4}
12 & 24 & 56771 & 0.55440 \\
10 & 20 & 29091 & 0.28409 \\
8  & 16 & 7330  & 0.07158 \\
11 & 23 & 4489  & 0.04383 \\
18 & 40 & 2771  & 0.02706 \\
9  & 19 & 1638  & 0.01599 \\
22 & 52 & 122   & 0.00119 \\
20 & 48 & 63    & 0.00061 \\
14 & 30 & 38    & 0.00037 \\
21 & 49 & 33    & 0.00032 \\
24 & 56 & 24    & 0.00023 \\
23 & 53 & 19    & 0.00018 \\
17 & 37 & 11    & 0.00010 \\
  & & & \\
& & &\\
& & & \\  \hline \hline
\end{tabular}
\hspace{0.5cm}
\begin{tabular}{cc|cc}
\multicolumn{4}{ c}{Mixture no.  2} \\
\hline \hline
$Z$  & $A$  & $N$     & $n$       \\ \cline{1-4}
25 & 57 & 24788 & 0.24207 \\
24 & 55 & 23909 & 0.23348 \\
26 & 58 & 20671 & 0.20186 \\
22 & 52 & 20160 & 0.19687 \\
23 & 53 & 5770  & 0.05634 \\
18 & 40 & 1570  & 0.01533 \\
20 & 48 & 1362  & 0.01330 \\
12 & 26 & 1348  & 0.01316 \\
21 & 49 & 1218  & 0.01189 \\
8  & 16 & 678   & 0.00662 \\
10 & 21 & 349   & 0.00340 \\
11 & 25 & 327   & 0.00319 \\
28 & 62 & 190   & 0.00185 \\
17 & 38 & 37    & 0.00036 \\
14 & 31 & 16    & 0.00015 \\
15 & 34 & 7     & 0.00007 \\ \hline \hline
\end{tabular}
\hspace{0.5cm}
\begin{tabular}{cc|cc}
\multicolumn{4}{ c}{Mixture no.  3} \\
\hline \hline
$Z$  & $A$  & $N$     & $n$       \\ \cline{1-4}
26 & 59 & 43766 & 0.42740 \\
25 & 57 & 18638 & 0.18201 \\
24 & 55 & 13079 & 0.12772 \\
22 & 52 & 12154 & 0.11869 \\
28 & 62 & 5020  & 0.04902 \\
23 & 53 & 3166  & 0.03091 \\
27 & 63 & 3159  & 0.03084 \\
20 & 48 & 1076  & 0.01050 \\
21 & 49 & 839   & 0.00819 \\
18 & 40 & 767   & 0.00749 \\
12 & 25 & 356   & 0.00347 \\
8  & 16 & 247   & 0.00241 \\
10 & 21 & 81    & 0.00079 \\
11 & 25 & 52    & 0.00050 \\ 
& & & \\
& & & \\ \hline \hline
\end{tabular}
%Line Break
\vspace{0.5cm}
%\hline \hline
%\newline
\begin{tabular}{cc|cc}
\vspace{0.1cm}\\
\multicolumn{4}{ c}{Mixture no.  4} \\
\hline \hline
$Z$  & $A$  & $N$     & $n$       \\ \cline{1-4}
28 & 65 & 60679 & 0.59256 \\
26 & 60 & 14103 & 0.13772 \\
27 & 63 & 9579  & 0.09354 \\
25 & 57 & 4088  & 0.03992 \\
24 & 55 & 3394  & 0.03314 \\
30 & 68 & 2906  & 0.02837 \\
22 & 52 & 2770  & 0.02705 \\
12 & 24 & 2215  & 0.02163 \\
29 & 69 & 726   & 0.00708 \\
23 & 53 & 724   & 0.00707 \\
20 & 48 & 369   & 0.00360 \\
18 & 40 & 240   & 0.00234 \\
21 & 49 & 208   & 0.00203 \\
10 & 20 & 188   & 0.00183 \\
8  & 16 & 172   & 0.00167 \\
14 & 30 & 39    & 0.00038 \\ 
& & & \\ \hline \hline
\end{tabular}
\hspace{0.5cm}
\begin{tabular}{cc|cc}
\vspace{0.1cm}\\
\multicolumn{4}{ c}{Mixture no.  5} \\
\hline \hline
$Z$  & $A$  & $N$     & $n$       \\ \cline{1-4}
28 & 65 & 37991 & 0.37100 \\
30 & 70 & 34071 & 0.33272 \\
29 & 69 & 11097 & 0.10836 \\
26 & 60 & 8322  & 0.08126 \\
27 & 63 & 3188  & 0.03113 \\
12 & 24 & 2060  & 0.02011 \\
31 & 73 & 1461  & 0.01426 \\
24 & 56 & 1308  & 0.01277 \\
25 & 57 & 1057  & 0.01032 \\
22 & 52 & 756   & 0.00738 \\
32 & 74 & 426   & 0.00416 \\
23 & 53 & 189   & 0.00184 \\
14 & 28 & 171   & 0.00166 \\
20 & 48 & 136   & 0.00132 \\
11 & 25 & 87    & 0.00084 \\
18 & 40 & 80    & 0.00078 \\ 
& & & \\ \hline \hline
\end{tabular}
\hspace{0.5cm}
\begin{tabular}{cc|cc}
\vspace{0.1cm}\\
\multicolumn{4}{ c}{Mixture no.  6} \\
\hline \hline
$Z$  & $A$  & $N$     & $n$       \\ \cline{1-4}
30 & 69  & 32165 & 0.31411 \\
28 & 64  & 27842 & 0.27189 \\
42 & 100 & 8896  & 0.08687 \\
32 & 76  & 5652  & 0.05519 \\
40 & 96  & 5418  & 0.05291 \\
38 & 90  & 4606  & 0.04498 \\
34 & 80  & 4140  & 0.04042 \\
36 & 84  & 3406  & 0.03326 \\
44 & 103 & 2602  & 0.02541 \\
41 & 99  & 1681  & 0.01641 \\
35 & 85  & 1388  & 0.01355 \\
26 & 60  & 1296  & 0.01265 \\
33 & 81  & 1088  & 0.01062 \\
31 & 73  & 850   & 0.00830 \\
39 & 93  & 714   & 0.00697 \\
16 & 36  & 499   & 0.00487 \\
17 & 39  & 157   & 0.00153 \\ \hline \hline
\end{tabular}
\vspace{0.3cm}
%\hline \hline
\end{table*}
}

\end{document}